\documentclass[12pt,prl,aps,floats,preprint]{revtex4-1}

\usepackage[german]{babel}
\usepackage[T1]{fontenc}
\usepackage[latin1]{inputenc}

\def\3{{\ss{}}}

\def\begineq{\begin{equation}}
\def\endeq{\end{equation}}
\def\be{\begin{equation}}
\def\ee{\end{equation}}

\usepackage{graphicx}
\usepackage{dcolumn}
\usepackage{bm}
\usepackage{subfigure}
\usepackage{color}
\usepackage[colorlinks,citecolor = blue, linkcolor=red,hyperindex,CJKbookmarks]{hyperref}

\begin{document}

\title{News \& Views:  \\ The distance rule in times of  Corona}
  
  \author{Detlef Lohse}
\email{d.lohse@utwente.nl}
\affiliation{Physics of Fluids, Max-Planck Center Twente for Complex Fluid Dynamics, 
MESA+ Research Institute and J.M. Burgers Centre for Fluid Dynamics, 
Department of Science and Technology, 
University of Twente,  P.O. Box 217, 7500 AE Enschede, The Netherland, \\
and Max Planck Institute for Dynamics and Self-Organization, Am Fassberg 17, 37077 G\"ottingen, Germany}

\date{\today}

\maketitle


The entire world is struggling with the Corona epidemic and its aftermath and is doing everything 
 to slow the spread of this deadly virus. A central role is played by the so-called 'distance rule': The distance between people 
 should not be less than one and a half, or better yet, two meters. But what is this rule based on?

The surprising truth is that it is based on a theory of viral infection by droplets from the 1930s, which is now long outdated, and that, as we now know, a distance of even five meters is not necessarily 'safe'.

The picture that William F. Wells developed at that time [1] in connection with the transmission of tuberculosis was the following: The drops produced by sneezing and coughing would have a wide size distribution and would fly out of the mouth and nose without much interaction between them. The small droplets would hardly be a problem because they would evaporate very quickly in the air and leave dry and therefore less dangerous aerosol particles behind, while the large droplets would behave ballistically. In this model, which is still used for risk assessment, the border between large and small is set quite arbitrarily at a droplet diameter of 5 - 10 $\mu$m.
For comparison: a virus has a typical diameter of 100 nm. (Bacteria such as the tubercle bacillus are more than ten times as large.) A drop with a diameter of 1 $\mu$m can therefore still contain 10 viruses at a virus concentration of 1\%. 
Under which circumstances this is sufficient for an infection depends on the virus and is still unknown for the corona virus.

In recent years, the work of Lydia Bourouiba (MIT) on the fluid dynamics of sneezing and coughing has clearly shown that the picture developed by Wells [1] for risk assessment of respiratory disease transmission is clearly inadequate. In her 'Insight' article in the Journal of the American Medical Association [2], Lydia Bourouiba elaborates what her results of the last years mean for the risk assessment of corona virus transmission.

The easiest way to get an idea of this is to watch one of her high-speed movies of the sneezing process [see
 the movies from Ref. [2] or others of her movies on YouTube], which she and her colleagues have recorded in recent years [3, 4], see also Fig. 1. This clearly shows that the range and lifetime of the cloud of tiny saliva and mucus droplets is much greater than assumed in Wells' model [1], namely up to 8 meters and up to 10 minutes, instead of one to two meters and less than one second.

The reason for this is that the droplets of saliva and mucus are expelled together with warm and humid air, which considerably delays their evaporation. In addition, they are ejected as a cloud, whereby they protect each other against evaporation, so to speak: The rate of evaporation is determined by the moisture gradient on the surface of the droplets. This is, of course, much smaller in a cloud of droplets where each of the droplets releases water vapor to the environment than for individual, isolated droplets. These two effects together can easily delay the evaporation of the small droplets by a factor of 1000 [2].
Their lifetime is therefore not determined by their size itself, but by the fluid dynamics of the water vapor concentration in the spray cloud around them [5, 6].

And another fluid dynamic effect plays an important role: sneezing or even exhaling can be seen as a short pulse of a turbulent jet, which on the one hand has a higher temperature and in particular a higher water vapour concentration than the surrounding air and is therefore lighter, but on the other hand is heavier than the surrounding air because of its charge of small and tiny saliva and mucus droplets.
These two effects can partly compensate each other. Important factors that determine the density of the cloud are the entrainment effect of the surrounding air and the rate at which the heavy droplets sink to the ground. Over time, this second effect can even make the turbulent droplet cloud lighter than its surroundings and rise [3]. Lydia Bourouiba and collaborators have already modeled these effects in 2014 with a simple hydrodynamic model (see the sketch in Fig. 2). The model is in good agreement with measured data on the range of droplet clouds of defined size [3].

If the sneezing or coughing takes place in closed rooms where convection effects caused by wind blowing the droplet cloud are hardly present, but instead possibly an air conditioning system, the situation becomes quite fatal: The warm turbulent cloud of small and tiny droplets of saliva 
and mucus rises and lands in the ventilation system of the air conditioning system, distributing it throughout the whole building. It is therefore not surprising that the level of contamination of passengers and crew on cruise ships on which COVID19 has broken out is so high. In fact, the coronavirus has been detected in  the ventilation system of a hospital [7], which supports Bourouiba's theory of virus spread through turbulent and warm droplet clouds [2].

So what are suitable countermeasures against the spread of the coronavirus? Obviously, it makes sense to minimize the release  of saliva and mucus droplets into the environment, for example by sneezing and coughing in the crook of your arm (whereby, by the way, everyone can
 immediately convince himself or herself that the turbulent jet is indeed warm). Exhaling through a mask, in which the droplets get caught, clearly is 
  also useful in closed buildings, to reduce the input of contaminated droplets into the surroundings. 
In particular, the knowledge that the range of the turbulent jets loaded with saliva and mucus droplets is not one to two meters, but eight meters and more, should lead to suitable protective measures for medical personnel (high-quality masks, possibly protective clothing). 
The ventilation systems in buildings with infected persons should also be adapted.

Many questions are still completely unresolved: For example, the existing theories for the fragmentation of liquids [8], which determine the so important initial size distribution of the droplets, are almost all based on Newtonian liquids. As anyone who has ever had a bad cold knows, this assumption is certainly not correct: Nasal and pharyngeal secretions are non-Newtonian. These viscoelastic properties play an important, generally inhibiting role in the fragmentation process [4], which strongly influences the droplet size distribution and thus the lifetime and range of the turbulent droplet cloud. This also affects  the requirements on permeability of protective masks.

The outbreak of the corona pandemic has made the urgency of answering the many open questions about the fluid physics of sneezing, coughing, speaking, singing, and even exhaling very clear. 
Aerosols and their life time in air do seem to play a central role here \cite{x}. 
And it also made clear how important basic research can suddenly become in a subfield of a discipline that was previously considered a niche.

\begin{figure}[htb]
\includegraphics[width=9.73cm]{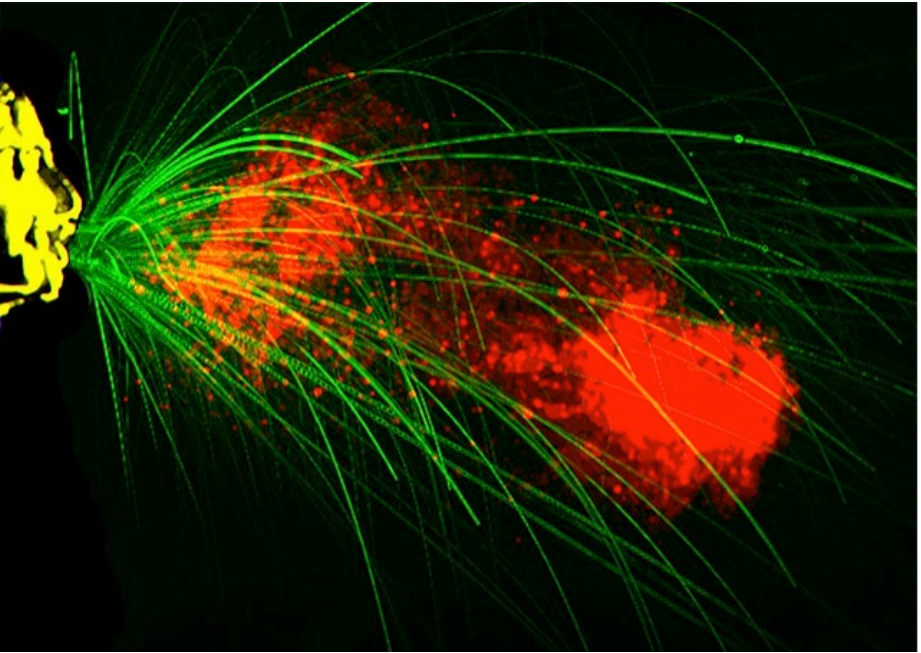}
\includegraphics[width=10cm]{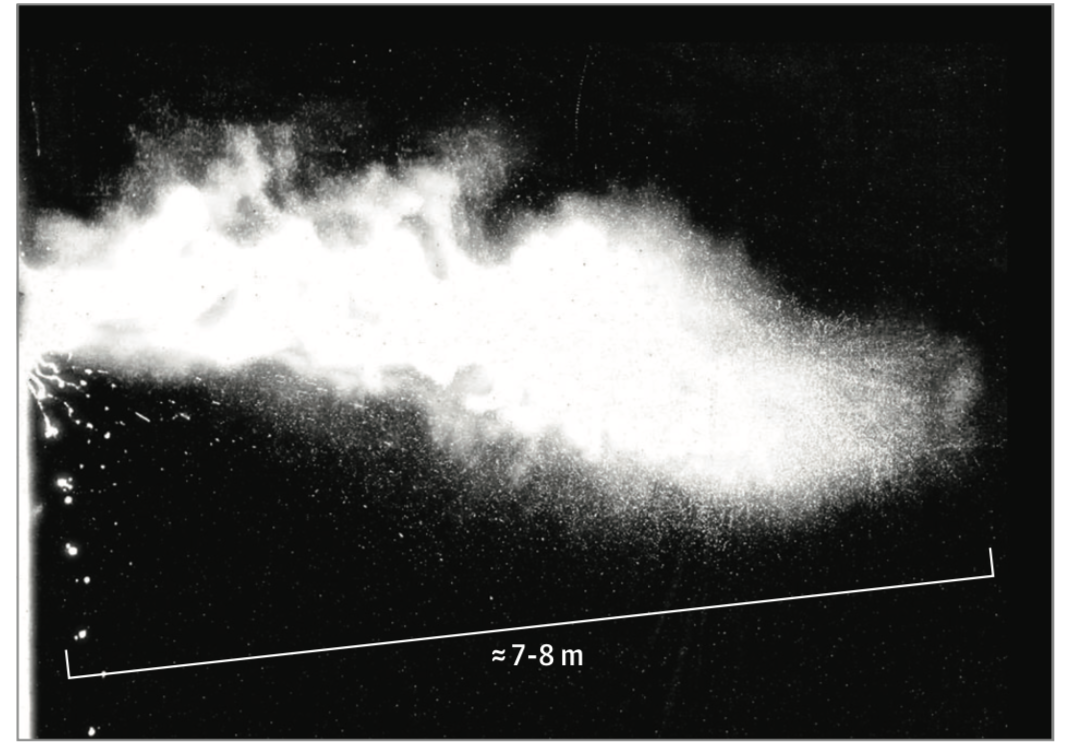}
    \caption{
    (a) High-speed imaging visualization of the sneezing process: On the left in yellow the head can be seen, also to get an idea of the length scale. On the one hand one 
     can see (in green) the trajectories of single large drops, which behave ballistically. On the other hand, 
      the turbulent jet of humid, warm air containing many small and tiny droplets of saliva and mucus is seen. 
      The cloud of these droplets can remain in the air for many minutes. The picture was taken in the group of Lydia Bourouiba (MIT) and was HHMI's Image of the Week: The Anatomy of a Sneeze [9]. (b) A visualization similar to (a), but now on a length scale of 8 m, which clearly shows the range of the turbulent droplet cloud of sneezing. From Ref. [2]. The corresponding movies for (a) and (b) can be seen [here].
}
\label{fig1}
\end{figure}

\begin{figure}[htb]
\includegraphics[width=15cm]{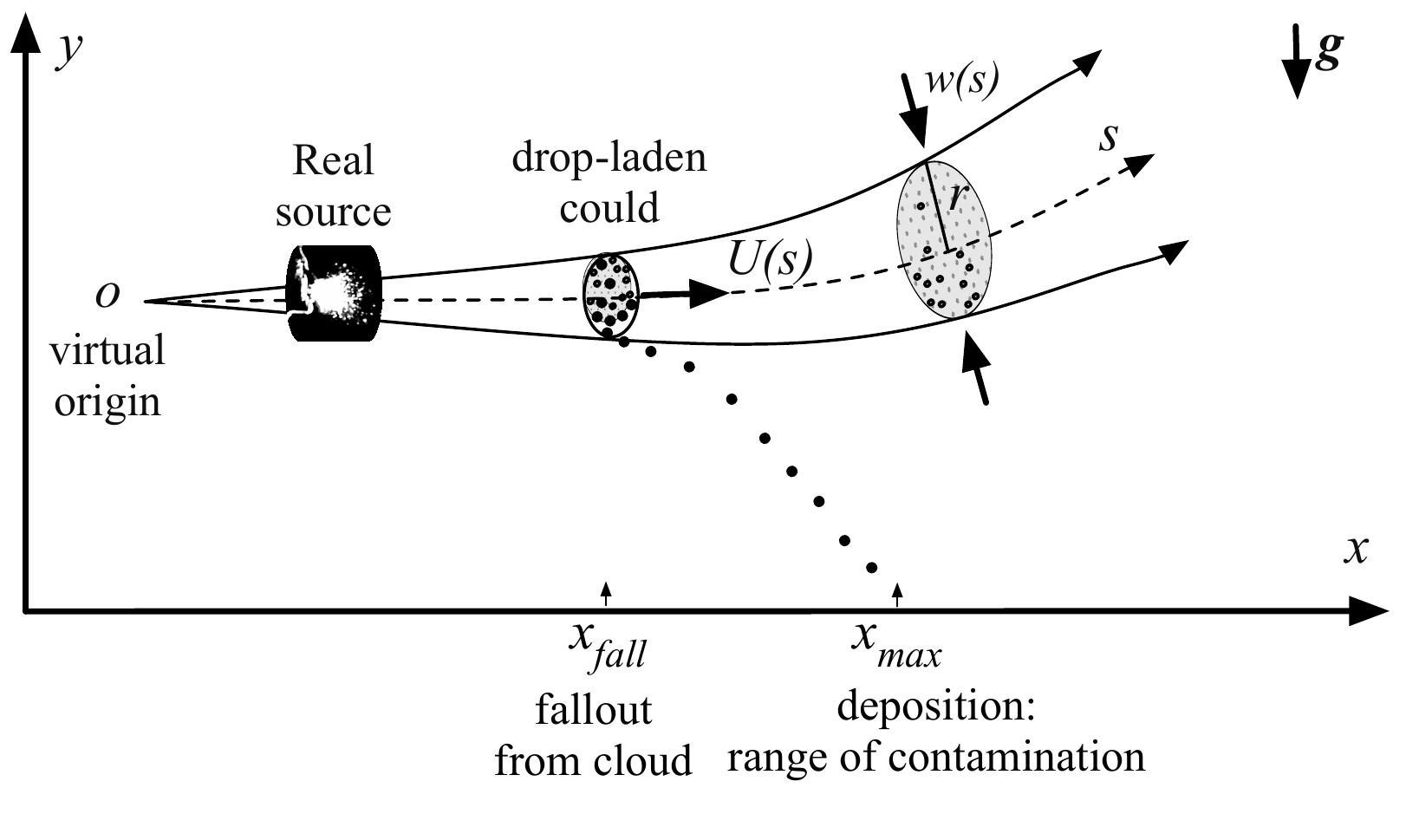}
    \caption{Schematic sketch to understand the dynamics of the warm, turbulent saliva and mucus droplet jet that occurs when sneezing or coughing [3]. Courtesy: Lydia Bourouiba, MIT.
    }
\label{fig2}
\end{figure}


\begin{thebibliography}{1}

\bibitem{wells1936}
W.~F. Wells and M.~W. Wells, {\em Air-borne infection}, J. Am. Med. Assoc. {\bf
  107},  1698  (1936).

\bibitem{bourouiba2020}
L. Bourouiba, {\em Turbulent Gas Clouds and Respiratory Pathogen Emissions:
  Potential Implications for Reducing Transmission of COVID-19}, J. Am. Med.
  Assoc. {\bf 323},  xxxx  (2020).

\bibitem{bourouiba2014}
L. Bourouiba, E. Dehandschoewercker, and J.~W. Bush, {\em Violent expiratory
  events: on coughing and sneezing}, J. Fluid Mech. {\bf 745},  537  (2014).

\bibitem{scharfman2016}
B. Scharfman, A. Techet, J. Bush, and L. Bourouiba, {\em Visualization of
  sneeze ejecta: steps of fluid fragmentation leading to respiratory droplets},
  Exp. in Fluids {\bf 57},  24  (2016).

\bibitem{derivas2016}
A. De~Rivas and E. Villermaux, {\em Dense spray evaporation as a mixing
  process}, Phys. Rev. Fluids {\bf 1},  014201  (2016).

\bibitem{villermaux2017}
E. Villermaux, A. Moutte, M. Amielh, and P. Meunier, {\em Fine structure of the
  vapor field in evaporating dense sprays}, Phys. Rev. Fluids {\bf 2},  074501
  (2017).



\bibitem{ong2020}
S.~W.~X. Ong, Y.~K. Tan, P.~Y. Chia, T.~H. Lee, O.~T. Ng, M.~S.~Y. Wong, and K.
  Marimuthu, {\em Air, surface environmental, and personal protective equipment
  contamination by severe acute respiratory syndrome coronavirus 2 (SARS-CoV-2)
  from a symptomatic patient}, J. Am. Med. Assoc. {\bf 323},  xxxx  (2020).

\bibitem{villermaux2007}
E. Villermaux, {\em Fragmentation}, Annu. Rev. Fluid Mech. {\bf 39},  419
  (2007).

\bibitem{poulain2019}
S. Poulain and L. Bourouiba, {\em Disease transmission via drops and bubbles},
  Phys. Today {\bf 72},  70  (2019).
  
  \bibitem{x}
  S.  Asadi, N.  Bouvier, A.  S. Wexler, and  W.  D. Ristenpart,
  {\em The coronavirus pandemic and aerosols: Does COVID-19 transmit via expiratory particles?},
  Aerosol Sci. Techn., April 2020; https://doi.org/10.1080/02786826.2020.1749229

\end{thebibliography}

\end{document}